\documentclass[12pt,a4paper]{article}

\usepackage[T1]{fontenc} 
\usepackage{amsmath}
\usepackage{amsfonts}
\usepackage{amssymb}
\usepackage{graphicx}
\usepackage{subfig}
\usepackage{hyperref}
\usepackage{color}
\usepackage{enumerate}

\newcommand{\dk}[1]{\frac{d^3#1}{(2\pi)^3}}
\newcommand{\Du}{\Delta_1}
\newcommand{\Dd}{\Delta_2}

\newcommand{\A}{{\cal A}}

\renewcommand{\O}{{\cal O}}
\newcommand{\be}{\begin{equation}}
\newcommand{\ee}{\end{equation}}
\newcommand{\beq}{\begin{equation}}
\newcommand{\eeq}{\end{equation}}
\newcommand{\bea}{\begin{eqnarray}}
\newcommand{\eea}{\end{eqnarray}}
\def\be{\begin{equation}}
\def\ee{\end{equation}}

\newcommand{\ep}{\epsilon}

\def\beq{\begin{equation}}
\def\eeq{\end{equation}}

\newcommand{\gam}{\gamma}

\newcommand{\zt}{\zeta}

\newcommand{\x}{\vec{x}}

\newcommand{\avg}[1]{\langle #1 \rangle}
\renewcommand{\k}{\vec{k}}
\newcommand{\q}{{\vec{q}}}

\newcommand{\prt}{\partial}

\newcommand{\Q}{\mathcal Q}

\newcommand{\ft}{f_{\rm NL}^\gam}
\newcommand{\tnl}{ \tau_{NL} }
\newcommand{\Quu}{Q_{11}}
\newcommand{\Qdd}{Q_{22}}
\newcommand{\Qud}{Q_{12}}
\newcommand{\Qut}{Q_{13}}
\newcommand{\Qdt}{Q_{23}}
\newcommand{\Qtt}{Q_{33}}
\newcommand{\expect}[1]{\left\langle #1 \right\rangle}

\begin{document}
\def\thefootnote{\fnsymbol{footnote}}

\begin{center}
\Large{\textbf{Tensor Squeezed Limits and the Higuchi Bound}} \\[0.5cm]
\end{center}
\vspace{0.5cm}

\begin{center}

\large{Lorenzo Bordin,$^{\rm a}$ Paolo Creminelli,$^{\rm b}$ Mehrdad Mirbabayi$^{\rm c}$ and  Jorge Nore\~na$^{\rm d}$}
\\[0.5cm]

\small{
\textit{$^{\rm a}$ SISSA, via Bonomea 265, 34136, Trieste, Italy}}

\vspace{.2cm}

\small{
\textit{$^{\rm b}$ Abdus Salam International Centre for Theoretical Physics\\ Strada Costiera 11, 34151, Trieste, Italy}}

\vspace{.2cm}

\small{
\textit{$^{\rm c}$ Institute for Advanced Study, Princeton, NJ 08540, USA}}

\vspace{.2cm}

\small{
\textit{$^{\rm d}$ Instituto de F\'isica, Pontificia Universidad Cat\'olica de Valpara\'iso, Casilla 4059, Valpara\'iso, Chile}}

\vspace{.2cm}

\end{center}

\vspace{.7cm}

\hrule \vspace{0.3cm}
\noindent \small{\textbf{Abstract}\\ 
We point out that tensor consistency relations---i.e.~the behavior of primordial correlation functions in the limit a tensor mode has a small momentum---are more universal than scalar consistency relations. They hold in the presence of multiple scalar fields and as long as anisotropies are diluted exponentially fast. When de Sitter isometries are approximately respected during inflation this is guaranteed by the Higuchi bound, which forbids the existence of light particles with spin: De Sitter space can support scalar hair but no curly hair. We discuss two indirect ways to look for the violation of tensor consistency relations in observations, as a signature of models in which inflation is not a strong isotropic attractor, such as solid inflation: (a) Graviton exchange contribution to the scalar four-point function; (b)  Quadrupolar anisotropy of the scalar power spectrum due to super-horizon tensor modes. This anisotropy has a well-defined statistics which can be distinguished from cases in which the background has a privileged direction.}

\vspace{0.3cm}
\noindent
\hrule
\def\thefootnote{\arabic{footnote}}
\setcounter{footnote}{0}

\section{Introduction}
The dynamics of spin-2 particles is theoretically very constrained. While General Relativity (GR) is the only consistent theory of an interacting massless spin-2 particle \cite{Weinberg:1964ew,Weinberg:1965rz}, there are tight theoretical constraints on the physics of a massive spin-2 \cite{deRham:2014zqa} and in general on modifications of GR. This theoretical robustness is particularly appealing and motivates the huge experimental effort dedicated to the study of gravitional waves (GWs) both of astrophysical and cosmological origin. The robustness of GR allows to predict in terms of few parameters the production of GWs by binary black holes. The same robustness shows up in the predictions for primordial tensor modes, which are way more model-independent than their scalar counterpart.  
For instance, the tensor power spectrum cannot be modified at leading order in derivatives \cite{Creminelli:2014wna, Kleban:2015daa}, at least in models which can be described within the framework of the Effective Field Theory of Inflation \cite{Cheung:2007st}.  

In this paper we explore another aspect of this robustness: the tensor consistency relations (CRs) \cite{Maldacena:2002vr}. Usually these relations are associated to single-field models \cite{Creminelli:2004yq} and, indeed, scalar CRs are in general violated when more than one field is relevant: a large $f_{\rm NL}^{\rm local}$ is generated in many multifield models and violates the CR for the 3-point function. On the other hand, it is easy to realize that tensor CRs still hold in multi-field models \cite{Dimastrogiovanni:2015pla}. As we will discuss in Section \ref{sec:argument}, the argument for which a long-wavelength GW can be locally removed by a suitable anisotropic change of coordinates (an adiabatic mode in the terminology of Weinberg \cite{Weinberg}) holds even in the presence of multiple scalar fields\footnote{Scalar CRs can also be seen as a consequence of the equivalence principle \cite{Creminelli:2013mca}. Notice however that scalar violation of the equivalence principle cannot spoil tensor CRs. }. More generally tensor CRs are violated only when there are light tensor perturbations which are not adiabatic, which means anisotropic perturbations are not efficiently damped. Therefore, while the violation of scalar CRs is a smoking gun of the presence of additional scalars, {\em the violation of tensor CRs would show that the Universe does not quickly evolve towards an anisotropic attractor during inflation}. (A similar conclusion about Solid Inflation was reached in \cite{Bartolo:2013msa, Akhshik:2014gja}.) The usefulness of such a signature becomes clear in view of the fact that at the level of background cosmology the isotropy of the observed universe puts an extremely weak constraint on the degree of anisotropy in the early universe. Background anisotropy rapidly dilutes during the thermal history. 

In the same way extra light scalars can violate scalar CRs, extra light spin-2 particles can violate tensor CRs by introducing long-lived anisotropies. Here, however, the theoretical constraints on spin-2 particles come into play. The Higuchi bound \cite{Higuchi:1986py}, as we will discuss in Section \ref{sec:Higuchi} and in the Appendices, forbids the existence of a spin-2 field in de Sitter (dS) space with a mass $m^2 < 2 H^2$, where $H$ is the Hubble constant of de Sitter. More generally, the Higuchi bound ensures that all perturbations with non-zero spin (and hence anisotropic) dilute faster than $\exp(-Ht)$. Therefore, although dS is allowed to have scalar hair, it cannot support curly hair. We will see that this is in some sense a stronger statement than Wald's no-hair theorem \cite{Wald} which assumes strong energy condition on matter fields: a condition that is violated by an innocuous light scalar field. Using the terminology of conformal field theory, we will discuss how primary composite operators are constrained by the Higuchi bound, while non-primary ones can evade it at the expense of introducing tachyonic instabilities. 

Since inflation occurs in a space-time which is approximately de Sitter, the bound should also apply to this case as long as de Sitter isometries are approximately respected by the active degrees of freedom. This ensures tensor CRs to hold in this subclass of models. Nevertheless, there are many inflationary models in which part of dS isometries are fully broken. Among those, tensor CRs are often violated in models with a symmetry pattern different from the one of the Effective Field Theory of Inflation, for example in Solid Inflation \cite{Gruzinov:2004ty, Golovnev:2008cf, Maleknejad:2011sq, Endlich:2012pz,Adshead:2013qp,Adshead:2013nka, Cannone:2014uqa,Bartolo:2015qvr}. In these cases tensor fluctuations, even on long wavelength, deform the background and are therefore not adiabatic. 

In Section \ref{sec:observations} we discuss the various observables sensitive to a violation of tensor CRs. The CMB correlator $\langle BTT \rangle $, recently studied in \cite{Meerburg:2016ecv}, is directly sensitive to the correlation function $\langle \gamma\zeta\zeta \rangle $. However, there is another slightly indirect probe of this correlation. In contrast to scalar entropy modes (such as extra light scalar fields), where the super-horizon fluctuations just redefine the observed homogeneous background in the observable Universe, non-adiabatic tensors leave a local imprint in the form of a quadrupolar anisotropy of the power spectrum. The anisotropy depends on the particular place in the Universe and we can only study its statistics: we find that the eigenvalues of the tensor which describes the anisotropy tend to be different. As such they are easy to distinguish from models with a preferred direction in the sky, which induce an axisymmetric quadrupolar modulation. 

Finally, the violation of tensor CRs shows up indirectly in the countercollinear limit of scalar correlation functions: the exchange of a light helicity-2 state generates a scalar 4-point function with a particular angular dependence, very different from the one usually parameterized by $\tau_{\rm NL}$. This 4-point function can be observed both in the CMB $\langle TTTT\rangle$ correlator or in future Large Scale Structure (LSS) surveys. Conclusions and future directions are discussed in Section \ref{sec:conclusions}.

\section{\label{sec:argument}Tensor consistency relations}

Squeezed limit CRs among cosmological correlation functions are in one to one correspondence with the adiabatic modes of Weinberg \cite{Weinberg}. Below we will give a brief derivation (details can be found in \cite{Hinterbichler,double}). Then we will focus on potential violations of tensor CR as an indicator of whether during inflation the background converges to an isotropic solution exponentially fast. 

Adiabatic modes in cosmology are super-horizon physical perturbations which are locally indistinguishable from a gauge mode. As such they do not affect short distance dynamics and therefore their correlation with the short distance perturbations is trivially related to a coordinate transformation. To find them Weinberg introduced the following trick

\begin{enumerate}[i]

\item First fix the gauge. Using ADM parametrization,
\be
ds^2 = -N^2 dt^2 + g_{ij}(dx^i +N^i dt)(dx^j +N^j dt),
\ee
the spatial part of the metric can be factorized as
\be
g_{ij} = a^2 e^{2\zeta} \left(e^{\gamma}\right)_{ij},\qquad \text{with} \quad \gamma_{ii} = 0.
\ee
The gauge can be fixed by imposing $\prt_i \gamma_{ij}=0$, and fixing time-reparametrization by choosing time-slices to coincide with constant energy density $\rho$ slices. This completely fixes the reparametrization freedom at finite wavelength. There are still asymptotic (non-vanishing at spatial infinity) spatial diffeomorphisms $x^i \to x^i+\xi^i(t,\x)$ which preserve the gauge condition. They satisfy \cite{Hinterbichler}
\be
\label{transverse}
\nabla^2\xi_i+\frac{1}{3}\prt_i\prt_j\xi^j=0.
\ee

\item Except for translations and rotations, applying these infinitesimal transformations to the FRW background excites linear metric perturbations:
\be
\label{adia0}
\zeta = \frac{1}{3}\prt_i\xi^i,\quad  N^i = \dot\xi^i,\quad
\gamma_{ij} = \prt_i\xi_j +\prt_j\xi_i - \frac{2}{3}\delta_{ij} \prt_k\xi^k,
\ee
where the dot denotes $d/dt$ and spatial indices are raised and lowered by $\delta_{ij}$ and its inverse. So one obtains a family of (trivial) infinite wavelength solutions to the equations of motion. 

\item The adiabatic modes are identified as the subfamily of solutions that can be deformed to finite wavelength. This requirement fixes the time-dependence of $\xi^i(t,\x)$.

\end{enumerate}

If adiabatic modes exist (that is if the last requirement can be satisfied), there is generically an infinite number of them. They can be organized by Taylor expanding the generating asymptotic diffeomorphisms at a fixed time-slice
\be\label{taylor}
\xi^{i}(t_0,\vec x) = \sum_{n=1}^\infty \frac{1}{n!} M_{i \,i_1\cdots i_n} x^{i_1}\cdots x^{i_n}.
\ee
The condition \eqref{transverse} translates into a trace condition on the matrices $M_{i \,i_1\cdots i_n}$. 

The $n^{th}$ order adiabatic mode in \eqref{taylor} leads to a CR that constrains the $\O(q^{n-1})$ term in the cosmological correlation functions with one soft momentum $\vec q \to 0$. Here we concentrate on the leading tensor CR for which
\be\label{xit}
\xi^i = \omega^i_j x^j, \qquad \omega^i_i =0,
\ee
under which $\gamma_{ij}\to \gamma_{ij} + 2\omega_{ij}$. If super-horizon tensor fluctuations become adiabatic, in the sense that to zeroth order in their wavenumber $q$ they can be locally removed by \eqref{xit} at all times, then the CR can be derived as follows. Violations of this condition will be discussed in Section \ref{sec:violate}. 

Consider the change of a short distance correlator under the transformation \eqref{xit}, $\delta_\omega \expect{O}$. For instance, $O$ can be the product of two scalar fluctuations at small separation compared to $q$: $\expect{\zeta(\vec x_1)\zeta(\vec x_2)}$ with $q|\vec x_1 -\vec x_2|\ll 1$ and 
\be\label{rhs}
\delta_\omega\expect{\zeta(\vec x_1)\zeta(\vec x_2)} = \expect{\zeta((\delta^i_j+\omega^i_j) x_1^j)\
\zeta((\delta^i_j+\omega^i_j) x_2^j)} - \expect{\zeta(\vec x_1)\zeta(\vec x_2)}.
\ee
If the short wavelength modes are in Bunch-Davies vacuum, which ensures that they are not excited until the mode $\q$ crosses the horizon, this can be related to the correlation function in the presence of a long wavelength tensor fluctuation:
\be\label{lhs}
\left.\delta_\omega\expect{O} = 2\omega_{ij} 
\frac{\delta}{\delta\gamma_{ij}}\expect{O}_\gamma \right|_{\gamma =0}
\simeq\lim_{q\to 0} \sum_s \ep_{ij}^s 
\omega_{ij}\frac{1}{P_\gamma(s,q)}
\expect{\gamma^s_\q O},
\ee
where we introduced the polarization vectors by expanding $\gamma^{ij}_\q = \sum_{s=1,2} \ep_{ij}^s \gamma^s_\q$, normalized to $\epsilon_{ij}^s \epsilon_{ij}^{s'}=2 \delta^{s s'}$. The tensor power spectrum is defined 
\be
\langle \gamma^s_{\vec q}\, \gamma^{s'}_{-\q}\rangle' = \delta^{s s'} P_\gamma(s,q),
\ee
where prime on the expectation value indicates that the momentum conserving delta function and a factor of $(2\pi)^3$ are removed. In writing \eqref{lhs} we have used the fact that super-horizon fluctuations of $\gamma_{ij}$ can be treated as a classical background up to corrections of order $q^3$ (see e.g. \cite{double} for a derivation). Taking $O$ to be the product of two scalar modes as in \eqref{rhs} and transforming them to momentum space, we obtain the following expression for the squeezed tensor-scalar-scalar correlation function \cite{Maldacena:2002vr}
\be
\lim_{\vec q \to 0}\frac{1}{P_\gamma(s,\q)} \langle\gamma^s_{\vec q} \;\zeta_{\vec k_1}\zeta_{\vec k_2}\rangle' 
= -  \epsilon^s_{ij} k_1^i k_1^j\frac{\partial}{\partial k_1^2}\langle \zeta_{\vec k_1}\zeta_{\vec k_2}  \rangle' 
\simeq \frac32  \epsilon^s_{ij} \hat k_1^i \hat k_1^j\langle \zeta_{\vec k_1}\zeta_{\vec k_2}  \rangle'  \;,
\ee
where in the last passage we neglected the small deviation from scale invariance of the scalar spectrum.

It is worthwhile stressing that in the derivation above we did not assume that inflation is single-field: the argument applies to cases with multiple scalars, like the curvaton scenario \cite{Lyth:2001nq} or quasi-single field models \cite{Chen:2009we}. We do not have to assume that the short-wavelength scalar perturbations will eventually become adiabatic: the operator $O$ above could include isocurvature perturbations. Also we did not have to assume that inflation is an attractor for scalar perturbations: the argument works for non-attractor models \cite{Namjoo:2012aa} as well. The tensor consistency relation is simply equivalent to the adiabaticity of tensor perturbations

Of course we never observe the $\q =0$ mode. So the above relation is useful because it applies at finite $q\ll k$ up to corrections of order $q$, as long as the vacuum state is Bunch-Davies. That is, the sub-horizon modes are not excited until their momentum redshifts to $k/a \sim H$. On the other hand, if the short wavelength modes are excited at a physical frequency $\omega > H$, then the above CR holds for sufficiently squeezed correlators: $q/k \ll H/\omega$. This ensures that the mode $q$ crosses the horizon and becomes adiabatic well before the short modes are excited \cite{Flauger}. Two concrete examples are: (1) Models with significant scalar and tensor emission from secondary sources (see for example \cite{Cook:2013xea, Mukohyama:2014gba, Ferreira:2014zia, Ferreira:2015omg}). Here one would expect most of the observed modes to be excited at $\omega \sim H$ because of the rapid expansion and redshift \cite{Mirbabayi:2014jqa}. (2) Inflationary models with periodic features in their potential \cite{Flauger_axion}, where $\omega\gg H$ is possible.  For all these models the tensor CRs hold for sufficiently small $q$.

\subsection{Violation of tensor CR and cosmic no-hair theorem}\label{sec:violate}

The Universe we observe is extremely isotropic at large scales. However, unlike the large scale homogeneity and flatness, the present isotropy implies an extremely weak constraint on the degree of anisotropy at the onset of the hot phase of cosmology. Anisotropies quickly dilute during the thermal history (and indeed we never asked inflation to solve the ``anisotropy problem'' !). Thus we cannot rule out inflationary models that are not strong (i.e.~exponential) isotropic attractors. However, the decay of anisotropies during the thermal history seems to imply that cosmological observables are insensitive to this possibility. 

This conclusion is not necessarily true. During inflation quantum fluctuations probe all possible small deformations of the background history, including anisotropic deformations, and the super-horizon correlation functions are a recorded memory of how those deformations evolve. In particular, suppose fluctuations of $\gamma_{ij}$ do not become adiabatic exponentially fast after horizon crossing. Then local observers would be able to detect the effect of the super-horizon $\gamma_{ij}$. They experience living in an anisotropic quasi-de Sitter Universe, which to their surprise does not isotropise exponentially fast. 

As discussed in the previous section, a violation of tensor CRs implies the existence of light non-adiabatic tensor perturbations: in this case inflation is not a strong isotropic attractor, or in other words, it supports anisotropic hair. This appears to be in contradiction to Wald's no-hair theorem, which states that homogeneous cosmologies with cosmological constant (CC) and any type of matter that satisfies strong and dominant energy conditions (respectively SEC and DEC) approach isotropic de Sitter space exponentially fast \cite{Wald}. However, the assumption of SEC is too strong to be of interest in the present discussion. Let us denote the components of the stress-energy tensor of the perturbations by an index $X$. When the back-reaction of perturbations on geometry is negligible DEC ($\rho_X >0$) and SEC ($\rho_X +3p_X>0$) imply
\be
\dot \rho_X = - 3 H(\rho_X +p_X) <-2 H \rho_X,
\ee
which implies an exponentially fast decay $\rho_X\propto a^{-2}$ of the underlying perturbations. However, even the superhorizon fluctuations of a light scalar field $\sigma$ violate SEC, because
\be
\rho_X+3p_X = 2 \dot\sigma^2 -  \tilde m_\sigma^2 \sigma^2 
\simeq - \tilde m_\sigma^2 \sigma^2<0,
\ee
where we used the fact that in the limit $\tilde m^2 = m_\sigma^2+2H^2 \ll H^2$ the time derivative behaves as 
\be
\dot\sigma = -\Delta_\sigma H \sigma,\qquad \Delta_\sigma \simeq \frac{\tilde m_\sigma^2}{3H^2}
\ee
and hence the kinetic term is negligible. (Note that $m_\sigma$ is the mass of a conformally coupled field. We use this for consistency with the following discussion about particles with spin.) In this model the deviation from dS,
\be
\rho + p = \dot\sigma^2 = \Delta_\sigma^2 H^2 \sigma^2,
\ee
decays arbitrarily slowly for sufficiently small $\tilde m_\sigma^2>0$. The fact that in the presence of a light scalar field dS is not an exponential attractor is the perturbative manifestation of why slow-roll inflation can occur. This shows that both de Sitter space and inflation can have scalar hair. But can they support anisotropic hair? 

A necessary condition for a no-hair theorem to exist is for it to hold for small perturbations. To study them it is useful to distinguish two qualitatively different cases

\begin{enumerate}[I]

\item {\bf dS isometries are approximately respected:} In this case the degrees of freedom furnish the representation of dS isometry group, which at super-horizon scales coincide with those of a $3d$ conformal field theory. This is an especially interesting case because the limit of exact dS is continuous. A long lived tensor degree of freedom (and in general any degree of freedom with nonzero spin) except for an adiabatic $\gamma_{ij}$ would constitute an anisotropic hair. However, as we will discuss in detail in Section \ref{sec:Higuchi}, all such degrees of freedom are either forbidden by the Higuchi bound, if they are fundamental fields, or they are composite operators made of tachyonic scalar fields. So at least perturbatively there exists a no-hair theorem for particles with spin: we can say {\em de Sitter does not have curly hair} \footnote{Since an adiabatic tensor mode is not locally observable, for us it does not constitute a genuine hair.}. 

\item {\bf Some dS isometries are fully broken:} Inflation has a preferred time, so dS isometries don't have to be respected by perturbations. We do not attempt to classify all possibilities but only list some of the known examples:

\begin{enumerate}[a]

\item {\bf Effective Field Theory (EFT) of Inflation:} applicable when there is no preferred spatial frame \cite{Cheung:2007st}. 
Scalar fields cannot have anisotropic stress at super-horizon scales \cite{Maleknejad} so that tensor modes become adiabatic. However, it might be possible to consider vector or tensor degrees of freedom which evade dS bounds: while light particles with spin are pathological in de Sitter (see Section \ref{sec:Higuchi}), one can imagine that the coupling with the preferred foliation makes them healthy. For instance it is well known that one can change the 2-point function of vector perturbations adding a suitable function of the inflaton in front of the vector kinetic term $f(\phi) F^2$  \cite{Ratra:1991bn, Watanabe:2009ct}: in this case the correlation can decay slower than the Higuchi bound outside the horizon.

\item {\bf Preferred spatial frame:} which exists when there is space-dependent background fields. All existing examples of violation of tensor CR like Solid Inflation \cite{Endlich:2012pz,Endlich:2013jia} (see also \cite{Gruzinov:2004ty}), Gauge-flation \cite{Maleknejad:2011sq} or Cromo-natural inflation \cite{Adshead:2013qp,Adshead:2013nka} fall in this category. As illustrated below in the example of solid inflation the long wavelength $\gamma_{ij}$ are not adiabatic. Moreover, additional tensor degrees of freedom often arise in these scenarios and they are not constrained by dS symmetries because of the presence of the additional background fields. \footnote{Another example in which the dS isometries are broken is massive gravity when the fiducial metric is not the inflationary dS: in this case one has an additional background with different symmetries and the Higuchi bound does not apply straightforwardly (for a recent discussion in the context of the ghost-free massive gravity see \cite{Fasiello:2012rw}).} \footnote{It is worth noting that even in the latter case where there is long-lived or growing anisotropy during inflation the anisotropic expansion rate cannot exceed $\O(\ep H)$ \cite{Maleknejad}.}

\end{enumerate}
\end{enumerate}

As a concrete example consider Solid Inflation. The degrees of freedom consist of perturbations of the Lagrangian coordinates $\{\phi^i\}$ of the solid
\be
\pi^i = \phi^i -x^i
\ee
and transverse gravitons $\gamma_{ij}$. The observed scalar modes are related to $\pi^i$ fields through $\zeta =\frac{1}{3} \prt_i\pi^i+\O(\pi^2)$. The super-horizon fluctuations of $\gamma_{ij}$ are not adiabatic in this model and tensor CR are violated. Indeed an anisotropic rescaling \eqref{xit} excites 
\be
\delta_\omega \pi^i = \omega^i_j x^j,
\ee
and $\prt_i\pi_j$ is a locally observable quantity. It leads to a long lasting super-horizon anisotropic stress. This allows us to relate the violation of tensor CR to the squeezed limit of scalar correlation functions. Let us write the correlation of two short modes in the presence of long tensor $\gamma_{ij}$ and scalar $\pi^i$ fluctuations as \cite{squeezed}
\be
\expect{\zeta(\x_1)\zeta(\x_2)}_{\gamma,\pi}= \xi(r) + \gamma_{ij}\xi^{ij}_\gamma(\vec r) 
+ \prt_i\pi_j \xi^{ij}_\pi(\vec r)+\cdots
\ee
where $\vec r = \x_2 -\x_1$. Under an anisotropic rescaling that removes $\gamma_{ij}$, $\prt_i\pi_j\to \prt_i\pi_j+\frac{1}{2}\gamma_{ij}$. Thus we get the following correction to the tensor CR
\be
\lim_{q\to 0} \frac{1}{P_\gamma(q)}\expect{\gamma_\q^s \zeta(\x_1)\zeta(\x_2)}-{\rm CR} 
= \frac12 \ep^s_{ij}\xi_\pi^{ij}(\vec r).
\ee
The same function $\xi_\pi^{ij}$ determines the squeezed limit of $\expect{\zeta_\q \zeta_{\k_1}\zeta_{\k_2}}$ and the counter-collinear limit of the scalar trispectrum $\expect{\zeta_{\k_1}\zeta_{\k_2}\zeta_{\k_3}\zeta_{\k_4}}$, when $\k_1 +\k_2 = -\k_3-\k_4 =\q$ is much smaller in magnitude than $k_1$ and $k_3$. One can check that this is indeed the case in the explicit calculations of the squeezed limit of \cite{Endlich:2013jia}. If $\xi_\pi^{ij}$ has a traceless part then the quantum fluctuations of $\pi^i$ fields also induce anisotropies in the local statistics of short modes. These anisotropies are larger by the scalar to tensor ratio $r^{-1}$. The observation imprints of such anisotropies will be studied in Section \ref{sec:observations}.

\section{\label{sec:Higuchi}Exact-de Sitter limit and the Higuchi bound}

As mentioned in the previous Section to have an isotropic attractor, all degrees of freedom with nonzero spin except for the graviton have to decay exponentially outside the horizon. This guarantees that the tensor CRs hold. The situation is analogous to scalar CRs which could be violated in the presence of additional light scalar degrees of freedom, also known as entropy perturbations. Different patches of the Universe with different entropy fluctuations experience different histories. If the entropy fluctuations mix with the adiabatic fluctuations during the cosmic evolution the absence of a unique history leads to a violation of scalar CRs.

Non-vanishing scalar fluctuations do not lead to anisotropy. In this Section we will review the Higuchi bound and show that non-pathological higher-spin fields indeed decay as $a^{-1}$ or faster in dS. Our focus will be on spin-1 fields and the more important case of spin-2 fields where the Higuchi bound forbids the mass range $0 < m^2 < 2H^2$ \cite{Higuchi:1986py}. 

The pathology of long-lived spin-1 and spin-2 degrees of freedom can be seen in the 2-point function of the fields, which is fixed by the de Sitter symmetries \cite{Arkani-Hamed:2015bza}. De Sitter is a maximally symmetric space with 10 isometries. Apart from spatial translations and rotations, it is invariant under the following two transformations
\begin{eqnarray}
 \label{dilations} D &=& -\! i\, (\eta \partial_\eta +x^i  \partial_i) \\
 \label{SCT} K_i &=& 2 i\, x_i( \eta\partial_\eta + \vec x \partial_{\vec x}) +i(\eta^2 - |\vec x|^2)\partial_i \;.
\end{eqnarray}
In de Sitter, an elementary field $\phi$ with mass $m$ and spin $s$ has two eigenmodes which at late times go as powers of the conformal time $\phi_\pm \sim \eta^{\Delta_\pm}$, where $\Delta_\pm$ are given by
\begin{equation}
\Delta_{\pm} = \frac{3}{2} \pm \sqrt{\left(s - \frac{1}{2}\right)^2 - \frac{m^2}{H^2}}\,.
\label{delta}
\end{equation}
If $\Delta$ is real, the solution $\Delta_{-}$ dominates at late times. So  $D$ and $K_i$ act on the fields (taking $\Delta = \Delta_-$) as
\begin{equation}
D \to -i(\Delta +x^i \partial_i)\;, \qquad K_i \to -i(2 \Delta  x_i + 2  x_i \, ( x^i \partial_i) -|\vec x|^2 \partial_i) \,.
\end{equation}
These transformations are the same as conformal transformations in 3 spatial dimensions, with the conformal dimension of the fields determined by their late-time behavior \eqref{delta}, which is fixed by the mass and spin. 

The universality of dS results then follows from the fact that the 2-point correlation functions of primary fields in a conformal field theory are fixed by the symmetries, up to the overall normalization. The 2-point function of a spin-1 field $A_i$ takes the form \cite{Arkani-Hamed:2015bza}
\be
\label{A2real}
\left\langle{A^{i}(\vec{x}) A^{j}(0)}\right\rangle\propto\frac{1}{|\vec{x}|^{2\Delta}}
(\delta^{ij}-2\hat{x}^i\hat{x}^j), \qquad\text{with} \quad \hat x \equiv \frac{\vec x}{|\vec x|}\,,
\ee
and that of a spin-2 field $S^{ij}$
\be
\label{S2real}
\left\langle{S^{ij}(\vec{x}) S^{kl}(0)}\right\rangle\propto\frac{1}{|\vec{x}|^{2\Delta}}
(\delta^{ik}-2\hat{x}^i\hat{x}^k)(\delta^{jl}-2\hat{x}^j\hat{x}^l)+(k\leftrightarrow l) \,.
\ee
Going to Fourier space one gets
\be 
\langle \epsilon.A_{\vec k} \, \tilde \epsilon . A_{\!-\!\vec k} \rangle' \propto e^{ i \psi} + 2\frac{(2\!-\!\Delta)}{(\Delta\!-\!1)} +  e^{- i \psi}\,,
\ee
\begin{equation}\label{spin_2_2-point_fourier}
\langle \epsilon^2.S_{\vec k} \, \tilde \epsilon^2 . S_{\!-\!\vec k} \rangle' \propto e^{2 i \psi} + 4\frac{3\!-\!\Delta}{\Delta} e^{i \psi} +6\frac{(3\!-\!\Delta)(2\!-\!\Delta)}{(\Delta\!-\!1)\Delta} + 4\frac{3\!-\!\Delta}{\Delta} e^{-i \psi} + e^{-2 i \psi}\,,
\end{equation}
where $\vec{\epsilon}$ and $\vec{\tilde{\epsilon}}$ are polarization vectors which (following \cite{Arkani-Hamed:2015bza}) are chosen to be
\begin{equation}\label{null_vectors}
\vec{\epsilon} = (\cos\psi, \sin\psi, i)\,,\qquad\vec{\tilde{\epsilon}} = (1,0,-i)\,\qquad \text{for}\quad
\k =(0,0,k).
\end{equation}
When $\Delta$ goes below 1 (corresponding to $m^2<0$ for $s=1$ and $m^2 < 2 H^2$ for $s=2$), the helicity-0 component becomes negative: it becomes a ghost.\footnote{The singularity at the threshold is not necessarily a pathology. It signifies an enhanced gauge symmetry which renders the longitudinal mode non-dynamical. Similarly, the case $\Delta = 0$, $s=2$ is an exception, since it corresponds to a massless particle for which only the helicity-2 components are physical, the others being only gauge artifacts.} For fields of higher spin $s$ the Higuchi bound is always at $m^2 = s(s-1) H^2$ corresponding to $\Delta_- =1$. Thus at a perturbative level all anisotropic hair decay at least as $a^{-1}$, and as a consequence geometric anisotropies decay as $a^{-2}$.

As we saw, the bound is simply a consequence of dS invariance: in particular, it does not require that the spin-2 state is described by an effective field theory with a parametric separation between the mass and the cutoff. For example it applies to the tower of Kaluza-Klein gravitons. One can evade the bound considering departures from exact de Sitter invariance given that during inflation the metric is not exactly de Sitter. However, by continuity one does not expect a significant change of the bound due to this.

The Higuchi bound comes from the relation among the different helicities, which is a consequence of the full dS isometry group. Inflation is usually associated with a preferred foliation of dS (the case IIa of the previous Section) and this breaks the conformal isometries of de Sitter. Only dilation invariance is approximately respected. In the absence of those there is no relation among helicities and no Higuchi bound. Therefore a particle with a large coupling with the preferred foliation can evade the Higuchi bound.

\subsection{Composite operators}

Phrased in these general terms, the Higuchi bound looks very powerful, since it looks one can apply it to any spin-2 operator, and not only to elementary spin-2 particles. For example it seems it applies also to a composite operator built out of scalars $\partial_i\phi\partial_j\phi -\frac13 (\partial\phi)^2 \delta_{ij}$. Actually this conclusion is too quick and it is straightforward to verify that this operator does {\em not} have a 2-point function of the form of eq.~\eqref{S2real}. Indeed eq.~\eqref{S2real} only applies to {\em primary} operators of a CFT and the operator $\partial_i\phi\partial_j\phi -\frac13 (\partial\phi)^2 \delta_{ij} $ is not a primary. At first, the distinction between primaries and descendants in de Sitter seems odd: the transformation properties of a field in de Sitter is fixed by its indeces, independently of whether it is the derivative of another field or not. Why should there be difference between the spatial part of a field $A_\mu$ and the one of 
 $\partial_\mu\phi$? The difference stems from the different time dependence of the time components, $A_0$ and $\partial_0 \phi$ respectively.
For $A_\mu$ all components will asymptotically behave in the same way: $A_\mu(\vec x, \eta) \sim \bar A_{\mu}(\vec x) \eta^\Delta$. In this case under a de Sitter isometry, $A_0$ does not affect the transformation of $A_i$ which behave like a CFT primary. On the other hand if $\partial_i \phi \propto \eta^\Delta$, then $\partial_0 \phi \propto \eta^{\Delta-1}$. Now the time component grows faster for $\eta \to 0$ and one cannot neglect, for $\eta \to 0$, the first term on the RHS of eq.~\eqref{SCT}. One can check that taking this into account, $\partial_i \phi$ transforms differently than $A_i$.  

In a CFT a generic operator is a sum of primaries and descendants. Thus $\partial_i\phi\partial_j\phi -\frac13 (\partial\phi)^2 \delta_{ij}$ can be made primary by adding suitable descendant fields. To find them one can impose that the variation under a special conformal transformation vanishes at the origin: this is the definition of a primary field, while descendants change even at $x=0$.
In particular:
\be
\delta_K \partial_i \phi = \Delta b_i \phi + {\cal O}(x) \qquad \delta_K \partial_i\partial_j\phi = (\Delta+1)(b_i \partial_j+b_j\partial_i)\phi -\delta_{ij} b^k \partial_k \phi + {\cal O}(x) \;,
\ee
where $\Delta$ is the dimension of $\phi$. For the particular case at hand we find the following primary operator quadratic in $\phi$ (for simplicity we multiplied by $(2 \Delta +1)$ ):
\begin{eqnarray}
\label{O2}
S_{ij} & = & (2 \Delta +1 )(\partial_i\phi\partial_j\phi-\frac{1}{3}\delta_{ij}(\partial\phi)^2) - \Delta \left[\partial_i(\phi \partial_j\phi)- \frac13 \delta_{ij} \partial_k(\phi\partial_k\phi)\right]  \\ & =  & (\Delta+1) (\partial_i\phi\partial_j\phi-\frac{1}{3}\delta_{ij}(\partial\phi)^2) - \Delta(\phi\partial_i\partial_j\phi-\frac{1}{3}\delta_{ij}\phi\nabla^2\phi) \,.
\end{eqnarray}
We verify explicitly in Appendix \ref{S_ij_2-point} that the 2-point function of this operator is of the form \eqref{S2real} with dimension $\Delta_t = 2\Delta +2$. 

The operator $S_{ij}$ is now a primary and the Higuchi bound tells us that its longitudinal part should become ghost-like for sufficiently small $\Delta_t$. But how can this happen if we start from a scalar with a manifestly positive 2-point function in momentum space? The best way to understand what happens as we approach the bound is to think about the wavefunction of the Universe for $S_{ij}$. In the Gaussian approximation it is given by
\begin{equation}
\Psi[S_{ij}] \sim \mbox{exp}\left[ -\frac{1}{2} \int \!\! \frac{\mathrm{d}^3 k}{(2\pi)^3} \, S_{ij} S_{kl} \langle \sigma_{ij}( \vec k) \sigma_{kl}(\!-\!\vec k) \rangle'  \right]\,.
\end{equation}
Here $\sigma_{ij}$ represents the ``dual'' operator in the putative CFT dual. As a consequence of conformal invariance $\langle \sigma_{ij}( \vec k) \sigma_{kl}(\!-\!\vec k) \rangle$ has the same form as eq.~\eqref{spin_2_2-point_fourier}, but choosing the other branch of eq.~\eqref{delta}, i.e.~$\Delta = \Delta_+$ (this is checked explicitly in Appendix \ref{sec:waveuniverse}). In approaching the Higuchi bound $\Delta_- \to 1$ and thus $\Delta_+ \to 2$, we see that the wavefunction becomes broader and broader and it becomes non-normalizable at the Higuchi bound. In the explicit calculation in Fourier space of the spectrum of the composite operator, one always gets an IR divergence at the Higuchi bound as we verify explicitly in the Appendix \ref{S_ij_2-point} for the case of a spin-1 composite operator. This IR divergence is cut off by the first (and hence longest) modes that exit the horizon during inflation. However, the cutoff dependence breaks conformal symmetry of the correlator so the general form \eqref{spin_2_2-point_fourier} is no longer expected, neither is the negativity of the helicity-0 correlation function.\footnote{One must be careful with the real-space computations involving tachyons. Indeed if one uses the scalar 2-point function
\be\label{2pftachyon}
\left\langle\phi(x)\phi(0)\right\rangle = \frac{1}{|x|^{2\Delta}}\,, \qquad \Delta<0\,,
\ee
in the expression of the primary eq.~(\ref{O2}), one gets the real-space expression eq.~\eqref{S2real} even below the Higuchi bound. But again the appearance of ``negative probabilities'' is fictitious. In fact the usual quantization in momentum space guarantees that the momentum space correlators are positive definite. The problem with tachyons is that the Fourier transform from momentum to real space,
\be
\xi(\x) = \int d^3\k k^{2\Delta-3} e^{i\k\cdot\x},
\ee
is IR divergent and hence the real space correlator is {\em not} \eqref{2pftachyon}. Indeed eq.~\eqref{2pftachyon} cannot come from a positive Fourier-space spectrum, since 
$\xi(\x =0) = \int d^3\k P(k)$ must be positive while \eqref{2pftachyon} vanishes at coincidence point for $\Delta <0$. The correct 2-point function contains an IR divergent constant which physically describes the growth of the tachyon field in an eternal de Sitter.}

In summary, the conformal symmetry relates the various helicities as in eq.~\eqref{spin_2_2-point_fourier}, so that the pathology of the helicity-0 part becomes a pathology of the full operator. However we saw that IR divergences modify eq.~\eqref{spin_2_2-point_fourier} in the case of composite operators. Moreover, the contributions of descendants will change the ratio among the different helicities and in particular one can have a non-primary spin-2 operator with an arbitrarily small $\Delta_t$. However, such composite operators are made of tachyonic primary fields with negative dimension which grow exponentially fast at super-horizon scales. For example, for the operator $\partial_i\phi\partial_j\phi -\frac13 (\partial\phi)^2 \delta_{ij} $, we need the scalars to have $\Delta \simeq -1$ for the composite operator to have $\Delta_t$ close to zero. Thus long lived anisotropic hair can be obtained at the expense destabilizing dS by growing scalar hair.

\section{\label{sec:observations}Observational prospects} 
Tensor modes with a wavelength much longer than the present Hubble radius are unobservable if the consistency relation holds \cite{Pajer:2013ana}. In this case, the only observable effects arise when the tensor enters the Hubble radius: it induces tides which result in a quadrupolar modulation of the density field \cite{Masui:2010cz, Giddings:2011zd, Schmidt:2012nw, Dai:2013kra, Schmidt:2013gwa}. In the following we are going to neglect this ``standard'' effect, since we are interested in possible {\em additional} signatures due to the violation of the consistency relation.

We parametrize the squeezed limit of $\langle\gamma\zeta\zeta\rangle$ as 
\be\label{bispectrum_parametrization}
\avg{\gam^s_{\vec q}\, \zt_{\vec k_1}\zt_{-\vec k_1}} ' \sim \left(\ft+\frac32\right)\, \avg{\gam^s_{\vec q}\gam^{s}_{-\vec q}}' \, \langle \zeta_{\vec k_1}\zeta_{-\vec k_1}  \rangle'   \, \epsilon^s_{ij}(\vec q) \hat k_{1,i}\hat k_{1,j} \;.
\ee
In this way $\ft$ parametrizes deviations from the consistency relation: as we discussed, when the tensor is way out of the Hubble radius the only physical effects are $\propto \ft$.

The tensor-scalar-scalar correlation function can be directly tested by measuring the correlation between temperature and B-mode polarization in the CMB, $\avg{BTT}$ \cite{Meerburg:2016ecv}. A very rough estimation of the signal-to-noise of such three-point function, when noise is dominated by cosmic variance, is 
\bea
(S/N)^2 &=& \Omega \int\!\! \frac{d^2l_B}{(2\pi)^2} \frac{d^2l_T}{(2\pi)^2} \frac{{\avg{B_{l_B}T_{l_T}T_{l_T}}'\,}^2}{\avg{B_{l_B} B_{l_B}}' \, {\avg{T_{l_T} T_{l_T}}'\,}^2} \nonumber \\
&\simeq&  {\ft}^2 r \, \A_s \left( \frac{l_{T,\,max}}{l_{T,\,min}} \right)^2 \log \left( \frac{l_{B\,max}}{l_{B\,min}} \right) \, , 
\eea 
where the angular size of the survey and $l_{T,min}$ are related as $\Omega \simeq \frac{(2\pi)^2}{l^2_{T,\,min}}$. $\A_s$ is the amplitude of scalar power spectrum, $\A_s = 2.2 \times 10^{-9}$ \cite{Ade:2015xua}. We see that the signal is proportional to the combination ${\ft}^2\,r$. Note also that this depends on the range of scales over which B-modes are observed. The reader may refer to \cite{Meerburg:2016ecv} for a detailed analysis of the signal-to-noise of the $\avg{\gamma\zeta\zeta}$ 3-point function. 
The authors find that a futuristic experiment with $l_{T\,max} \sim 4500$ and $l_{B\,max} \sim 500$ should be able to reach  \footnote{Notice that ref.~\cite{Meerburg:2016ecv} uses a different notation compared to ours: $f_{\rm NL} ^{here}\, \sqrt r = 24 f_{\rm NL}^{there} $.} 
\be
\label{Bbound}
{\ft}^2 r \lesssim 6 \times 10^3 \;;
\ee
an experiment of this kind will be cosmic variance limited unless $r \lesssim 10^{-3}$.

Even if the tensor modes are not directly measured, the violation of the tensor CR can be observed looking at the statistics of scalar perturbations only. We are now going to study the effect on the scalar 2-point function and 4-point function. 

\subsection{Modulation of the scalar 2-point function}

As discussed above, if the consistency relation is violated, the effect of a super-horizon tensor mode is physical and can be observed locally (see for example \cite{Brahma:2013rua, Dimastrogiovanni:2014ina, Akhshik:2014bla,Emami:2015uva}). Here we focus on the modulation it induces on the scalar 2-point function\footnote{Superhorizon tensor modes also induce short-scale correlation among scalar and tensor perturbations (see e.g.~\cite{Akhshik:2014gja}).}
\be
\avg{\zt_{\vec k_1}\zt_{\vec k_2}}_{\gam} \simeq \avg{\zt_{\vec k_1}\zt_{\vec k_2}} + \gam^s_{\vec q} \ \frac{\avg{\gam^s_{\vec q}\,\zt_{\vec k_1}\zt_{\vec k_2}}'}{P_\gam(q)} \;,
\ee
which would be observed as a quadrupole in the power spectrum proportional to the average amplitude of the super-horizon spin-2 modes
\be
P_\zt(\vec k) = P_\zt(k)\left[ 1 + \ft \epsilon^s_{ij}(\vec q) \hat k_i \hat k_j \gam^s_q \right] \equiv P_\zt(k)\left[ 1 + \Q_{ij} \hat k_i \hat k_j \right]  \;.
\ee
The (squared) amplitude of $\Q_{ij}$ is obtained averaging over all the super-horizon modes:
\bea\label{quadrupole_amplitude}
\Q^2 &=& \frac{8\pi}{15} \avg{\Q_{ij}\Q_{ij}} 
=  \frac{8}{15\pi^2}\ {\ft}^2 \int_{q<H_0} dq \ q^2  \avg{\gam^s_{\vec q}\gam^{s}_{-\vec q}}' 
\approx \frac{4 }{15 }{\ft}^2  r \A_s \Delta N\, ;
\eea
where $r$ is the tensor-to-scalar ratio and $\Delta N$ is the number of e-folds of all modes outside the present Hubble radius.
Experimental limits come from the CMB: $\Q \lesssim 10^{-2}$  \cite{Ade:2015hxq}.\footnote{For comparison, notice that the quantity $g_2$ used in \cite{Ade:2015hxq} is given by: $g_2 \equiv \Q/\sqrt{5}$. } This gives:
\be
{\ft}^2 \ r =  \frac{75 }{4 } \  \frac{g_2^2}{\A_s} \frac{1}{ \Delta N} \lesssim 8.5\times 10^5 \frac{1}{ \Delta N} \;.
\ee
Notice that the measurable quantity is always the combination ${\ft}^2 r$, as in eq.~\eqref{Bbound}. In this case, since we do not observe the tensor mode directly, the measurement is also sensitive to additional helicity-2 states even if they are not correlated with gravitational waves. In this case the effect is proportional to the power spectrum of this extra state, instead of $r$, and its coupling with scalar perturbations, instead of $\ft$. 

Let us comment on the quantity $\Delta N$ which describes the cumulative effect of all super-horizon modes \cite{Bartolo:2014xfa}. If the tensor (or the additional helicity-2 field) has a (small) blue spectrum, $n_T >0$---this is the case of solid inflation---the integral converges in the IR. In this case, assuming inflation is sufficiently long, the sum over all modes gives $\Delta N \simeq n_T^{-1}$. For example, taking $n_T = 0.03$, the experimental bound above gives ${\ft}^2 \ r \lesssim 2.5 \times 10^4$. In models with a red spectrum, the integral is IR divergent and therefore the result depends on the duration of inflation. Notice also that if the tensor mass during inflation is negative, the model does not evolve towards isotropy but (slowly) away from it: in this case the initial condition for inflation must be carefully chosen to satisfy the experimental limits on ${\cal Q}$.

Future constraints on the amplitude $\Q^2$ will arise from the  new generation of experiments. For instance new LSS surveys may greatly improve the current limits.  In this case a very rough estimate of the signal to noise is given by
\bea
􏰈\left({S}/{N}\right)^2 &=& \frac{V}{2} \int 􏰐\dk k 􏰈\left(\frac{\Delta P􏰉(\vec k)}{P(k)}\right)^2  \nonumber \\
&\simeq&  {f_s} \frac{2}{45\pi} \A_s \left(\frac{k_{max}}{k_{min}}\right)^3 {\ft}^2 r \, \Delta N \;, 
\eea
where $f_s$ is the fraction of the sky covered by the survey. 
An improvement of $(k_{max}/k_{min})$ of order $10$ in the future experiments will put constraints on ${\ft}^2 r$ roughly of order $\mathcal O(100/\Delta N)$. Note, however, that LSS measurements of the quadrupole could be complicated by the fact that both gravitational non-linearities and redshift-space distortions induce a quadrupole. One can also look for this effect in the 21 cm power spectrum. A detailed analysis on the  bounds one could get on $\Q^2$ is given in  \cite{Shiraishi:2016omb}. 

\subsubsection*{The statistics of $Q_{ij}$}
\label{sec:statistics}

A quadrupolar modulation of the power spectrum is not only induced by tensor perturbations but also, in models like solid inflation, by scalars and vectors \cite{Bartolo:2014xfa}. The anisotropy in short-scale power spectrum depends on $(\partial_i\partial_j/\partial^2 -1) \zeta$  in the case of scalars and on $\partial_i/\sqrt{\partial^2} V_j$ in the case of a vector. Naively one may expect that the statistical distribution of the matrix $Q_{ij}$, i.e.~the distribution of its eigenvalues, depends on whether the origin of the quadrupolar modulation is a scalar, a vector or a tensor. 
Indeed for a single Fourier mode the matrix is quite different comparing scalars with tensors: for instance the scalar one has two equal eigenvalues and it is thus axially symmetric. However once we average over all nearly Gaussian super-horizon modes, this difference is lost: rotational invariance and Gaussianity imply that the distribution is uniquely fixed in terms of the variance
\be
\langle Q_{ij} Q_{kl}\rangle \propto \frac12 \left(\delta_{ik}\delta_{jl} + \delta_{il} \delta_{jk}\right) -\frac13 \delta_{ij}\delta_{kl} \;.
\ee
Since in all cases $Q_{ij}$ is Gaussian (it is always linear in an approximately free field), all correlation functions just reduce to the previous one and everything can be written in terms of the variance. Therefore there is no way to distinguish whether $Q_{ij}$ comes from long-wavelength fluctuations of a scalar, a vector or a tensor.

Let us now address a different question: whether one can distinguish a quadrupole generated by long-wavelength perturbations from the case in which the quadrupole is exactly axially symmetric \cite{Ackerman:2007nb}: $P_\zt(\vec k) = P_\zt(k)\left[ 1 + c (\hat k \cdot \hat n)^2 \right]$. This happens when there is a preferred direction $\hat n$ in the sky, for example due to the presence of a vector field in the background solution, see e.g.~\cite{Watanabe:2010fh, Bartolo:2012sd}. In this case two eigenvalues of the matrix $Q_{ij}$ are the same, while in general this will not hold for a $Q_{ij}$ with the Gaussian statistical distribution discussed above. If the observed $Q_{ij}$ is axially symmetric within the experimental uncertainties, this will disfavor a quadrupole generated by super-horizon fluctuations. We can easily quantify this statement and ask how well one should measure the quadrupole modulation before being able to rule out the Gaussian statistics discussed above. 

A similar problem arises when studying the statistics of the shear tensor in the Large Scale Structure \cite{Doroshkevich}. Given a symmetric traceless $3\times3$ matrix we want to find the probability density function (PDF) of the two independent eigenvalues. As we discussed, the entries of the matrix are Gaussian.  Since the distribution must be rotationally invariant it can only depend on $Q_{ij} Q_{ij}$:
\be\label{pdf_entries}
P(\Quu,\Qdd,\Qud,\Qut,\Qdt) = \mathcal{N}e^{-\frac{\Quu^2+\Qdd^2+\Qtt^2+2\Qud^2+2\Qut^2+2\Qdt^2}{2 \sigma^2}} d\Quu\,d\Qdd\,d\Qud\,d\Qut\,d\Qdt,
\ee  
where $\Qtt=-(\Quu+\Qdd)$ and $ \mathcal{N}$ a normalization constant. We want to perform a change of variables writing the PDF in terms of the eigenvalues of the matrix, respectively $a,\ b\ \mbox{and}\ c=-(a+b)$, plus the  three Euler angles, $\alpha,\ \beta,\ \gamma$. The Gaussian now depends on $a^2+b^2+c^2=2 (a^2+ab+b^2)$, while the Jacobian of the trasformation is
\be
|J|=(a-b)(a-c)(b-c) \sin \beta =(a-b)(2a+b)(a+2b)\sin \beta.
\ee
The eigenvalues are assumed to be in decrescent order, i.e. $a \geq b \geq c,$
this implies
$$ b \leq a,\ \ b \geq - a/2.$$
After integrating over the Euler angles we get (with a suitable change of the normalization  $\mathcal{N}$)
\be\label{eigenvalues_pdf}
P(a,b)= \mathcal{N} (a-b)(2a+b)(a+2b) e^{-\frac{a^2+ab+b^2}{\sigma^2}}. 
\ee
This PDF agrees with the one found by Doroshkevich \cite{Doroshkevich} after one sets Tr$(Q)=0$. Notice that the Jacobian suppresses the PDF when two eigenvalues are similar and this makes the distinction from the axially symmetric case easier.

We can now compute the probability of having two nearly degenerate eigenvalues. One has to integrate the PDF \eqref{eigenvalues_pdf} in the region 
$$ \left( \, \frac{|a-b|}{|a|} \leq \epsilon \ \bigcup \ \frac{|b-c|}{|b|}  \leq \epsilon \, \right) \ \ \bigcap\ \ \left( \, b \leq a,\ \ b \geq - \frac{a}{2} \, \right) \;.$$
The explicit integration gives 
\bea
P(\epsilon) &=& 1+ \frac{3\sqrt{3}}{4 \, \sqrt{(3-3\epsilon+\epsilon^2)^3}} - \frac{3\sqrt{3}}{4 \, \sqrt{3-3\epsilon+\epsilon^2}} + \frac{3\sqrt{3}}{4 \, \sqrt{(3+3\epsilon+\epsilon^2)^3}} - \frac{3\sqrt{3}}{4 \, \sqrt{3+3\epsilon+\epsilon^2}} \nonumber \\
&=& \frac{3}{8}\epsilon^2+\mathcal{O}(\epsilon^4).
\eea
The above formula tells us that the probability of having two eigenvalues that differ less than $10\%$ from each other is  $\sim 0.4\%$. Therefore if the errors on the quadrupolar modulation are reasonably small, one can rule out that $Q_{ij}$ is generated by many superhorizon Gaussian modes. Conversely, if the eigenvalues are observed to be different one can rule out all models with a preferred direction.

\subsection{The scalar 4-point function}

The exchange of a soft graviton gives a 4-point scalar correlator in the limit in which the sum of two momenta is small: $q\equiv | \vec k_1+ \vec k_2| \ll k_i$. One gets, neglecting the terms which respect the tensor CR,
\bea \label{trispectrum}
\avg{\zt_{\vec k_1}\zt_{\vec k_2}\zt_{\vec k_3}\zt_{\vec k_4}}' &\simeq& \big\langle \avg{\zt_{\vec k_1}\zt_{-\vec k_1}}\ \avg{\zt_{\vec k_3}\zt_{-\vec k_3}} \big\rangle  '\nonumber \\
&=& \sum_s \avg{\gam^s_{\vec q} \gam^s_{-\vec q}}' \ \frac{\avg{\gam^s_{\vec q}\,\zt_{\vec k_1}\zt_{-\vec k_1}}'}{P_\gam(q)} \ \frac{\avg{\gam^s_{\vec q}\,\zt_{\vec k_3}\zt_{-\vec k_3}}'}{P_\gam(q)}  \nonumber \\
&=& {\ft}^2 P_\gam(q) P_\zt( k_1)P_\zt(k_3) \sum_s \epsilon^s_{ij}(\vec q) \epsilon^s_{kl}(\vec q) \hat k_{1,i}\hat k_{1,j} \hat k_{3,k} \hat k_{3,l}\, .
\eea
The angular dependence of \eqref{trispectrum} can be cast in the form \cite{Seery:2008ax}
\be
\sum_s \epsilon^s_{ij}(\vec q) \epsilon^s_{kl}(\vec q) \hat k_{1,i} \hat k_{1,j} \hat k_{3,k} \hat k_{3,l}\equiv \cos2\chi_{12,34}\, ,
\ee
where $\chi_{12,34}\equiv \phi_1-\phi_3$ is the angle between the projection of $k_1$ and $k_3$ on the plane orthogonal to $q$. The final expression of the trispectrum due to a graviton exchange
%
%
is
\be\label{final_trispectrum}
\avg{\zt_{\vec k_1}\zt_{\vec k_2}\zt_{\vec k_3}\zt_{\vec k_4}}' = {{\ft}^2} P_\gam(q) P_\zt(k_1)P_\zt(k_3) \cos2\chi_{12,34}.
\ee
%

One can look for this kind of 4-point function directly in the CMB temperature map. The momentum dependence is similar to the standard $\tnl$ trispectrum shape
\cite{Byrnes:2006vq},
\be \label{trispectrum_parametrization}
\avg{\zt_{\vec k_1} \, \zt_{-\vec k_1+\vec q} \, \zt_{\vec k_3} \, \zt_{-\vec k_3-\vec q}}' = \frac{5}{3} {\tnl} P_\zt(q) P_\zt(k_1) P_\zt(k_3) \;.
\ee
Notice, however, that the additional angular dependence of eq.~\eqref{final_trispectrum} is such that the two shapes are effectively orthogonal and no constraint on eq.~\eqref{final_trispectrum} can be obtained from the $\tnl$ bounds: a dedicated analysis must be performed. It is easy to estimate what kind of constraints (or detection!) one should be able to get with a dedicated analysis. The bound on $\tnl < 2800 \ (95\% \,\mbox{CL})$ \cite{Ade:2013ydc} can be roughly converted in
\be
{\ft}^2 r \lesssim   4\times10^4 \;,
\ee
where we wrote $P_\gamma =  P_\zeta\, r/4$ and we took into account a factor of $\avg{\cos^2 \chi_{12,34}} = 1/2$ which arises when one integrates eq.~\eqref{final_trispectrum} over all configurations. It is important to stress that this bound is not so much worse than the futuristic limit using B-modes of eq.~\eqref{Bbound}. This strongly motivates a dedicated analysis of this shape of the 4-point function: a signal of tensors could be already in Planck temperature data, even without observing B-modes! Moreover this signal has the advantage of being sensitive to extra helicity-2 states which induce a 4-point function of the form \eqref{trispectrum_parametrization}, even in the absence of mixing  with gravitational waves.

One can also attempt to measure the scalar 4-point function by looking at galaxy number counts. By  correlating the amplitudes of a pair of 2-point functions, reference \cite{Jeong:2012df} studies whether future surveys will be able to observe this effect. It is straightforward to translate their results in our notation.
The variance of the effect induced by a violation of the consistency relation for future surveys is
\be
\left(\frac{S}{N}\right)^2 \simeq \frac{\pi}{6075} {{\ft}^4r^2} \A_s^2{\left(\frac{k_{max}}{k_{min}}\right)}^{6}\,.
\ee
For a survey like Euclid, for which $k_{min} \simeq 10^{-3}\, \mathrm{Mpc}^{-1}$, the expected constraints from this observable are ${\ft}^2r \lesssim 2\times 10^4 \, (0.1\,\mathrm{Mpc}^{-1}/k_{max})^3$, comparable to the limits that would be obtained by analyzing the CMB 4-point function. Note that in the event of a positive detection,
the above estimate of the signal-to-noise ratio in the 4-point function breaks down due to the non-Gaussian contribution to the noise. Beyond this point, an improved estimator (analogous to the one introduced in \cite{local}) is necessary to decrease the error-bars as $1/\sqrt{N_{\rm data}}$.

\section{\label{sec:conclusions}Conclusions and future directions} 
We showed that tensor CRs are very robust and that their violation would contain a lot of insight into the physics of inflation, showing that anisotropic perturbations are not quickly redshifted away. We know few explicit models which violate tensor CRs, but it would be nice in the future to study systematically this violation. In particular one could explore the connection between the general analysis of tensor mass terms done in \cite{Blas:2009my,Cannone:2014uqa} and the violation of tensor CRs. Since we are discussing non-adiabatic tensor perturbations, one would like to understand what happens at reheating and in particular if additional contributions which violate tensor CRs can arise at the end of inflation, similarly to what happens in the case of scalars. Since the Higuchi bound applies also to higher-spin states, much of what we said could be generalized to these cases as well. From the experimental side, it would be extremely interesting to get explicit constraints on the helicity-2 mediated scalar trispectrum, eq.~\eqref{final_trispectrum}, using Planck data and start thinking about future improvements from LSS data. We hope to come back to these issues in the near future.

\section*{Acknowledgements}
It is a pleasure to thank Peter Adshead, Atish Dabholkar, Guido D'Amico, Azadeh Maleknejad, Daan Meerburg, Marcello Musso, Leonardo Senatore, Marco Serone, Shahin Sheikh-Jabbari, Marko Simonovi\'c and Matias Zaldarriaga for useful discussions. M.M. is supported by NSF Grants PHY-1314311 and PHY-0855425. J.N. is supported by Proyecto VRIEA-PUCV 039.362/2016.

\appendix

\section{Composite spin-1 and spin-2 fields}
\label{S_ij_2-point}

\paragraph{Spin 1 in Fourier space}

Let us consider a primary spin $1$ operator, $\A_i$ with conformal weight $\Delta$. Its 2-point function is  
\be\label{vector_2-point}
\langle \epsilon . \A(\vec x) \, \tilde \epsilon . A(\vec 0) \rangle \propto \frac{ (\, \epsilon . \tilde \epsilon \,)-2(\epsilon . \hat x)(\tilde \epsilon . \hat x)} {| x|^{2\Delta}}.
\ee 
Fourier transforming the above expression we get  a power spectrum with the same pathology of the spin-2 case: the helicity zero mode gives a negative contribution if $\Delta<1$. Indeed,
\begin{equation}\label{vector_2-point_fourier}
\langle \epsilon . \A_{\vec k} \, \tilde \epsilon . \A_{\!-\!\vec k} \rangle' \propto k^{2\Delta-3} \left( e^{i \psi} + 2\frac{2-\Delta}{\Delta-1} + e^{-i\psi} \right).
\end{equation}
However, if we compute the power spectrum of a composite vector directly in Fourier space, it will be always positive. To see this concretely consider the composite operator,  
\begin{equation}
\A_i \equiv \Du \phi \partial_i \sigma - \Dd \, \sigma \partial_i \phi,
\end{equation} 
with $\phi$ and $\sigma$ two scalar fields with conformal weights, respectively $\Du$ and $\Dd$.  
After having verified that $\A_i$ satisfies Eq.~\eqref{vector_2-point} with weight $\Delta_v=\Du+\Dd+1$, we move to Fourier space.

In Fourier space $\A_i(\vec k)$ is a convolution of $\phi$ and $\sigma$:
\be
\A_i(\vec k) = \int \dk{p} i \left[ \Du \, \phi_{\vec p} \, (k\!-\!p)_i \, \sigma_{\vec k \!-\! \vec p}  - \Dd \, p_i \, \phi_{\vec p} \,   \sigma_{\vec k \!-\! \vec p} \right],
\ee
hence its power spectrum is given by
\begin{eqnarray}
\langle \epsilon.O_{\vec k} \, \tilde \epsilon. O_{\!-\!\vec k}\rangle \!\!\!\!\! &\propto&
\!\!\!\!\! \epsilon_i \tilde \epsilon_j  \!\! \int \!\! \dk p \frac{-\Du^{\,2} \, p_i \, p_j \! -\! \Du \Dd [ (p\!-\!k)_i \, p_j + p_i (p\! -\! k)_j ] \!+\! \Dd^{\,2}  (k\! -\!p)_i  (k\! -\! p)_j}{|\vec p - \vec k|^{3-2\Du} \ p^{3-2\Dd}}. \nonumber \\
\end{eqnarray}
The explicit expressions of $\epsilon$ and $\tilde \epsilon$ are given in Eq.(\ref{null_vectors}). We can split the two helicities  contributions in the two point function. Then we integrate over $\vec p$. We find that the helicity 1 contribution is IR divergent for $\Du<-1$ or $\Dd<-1$ while it is UV divergent for  $\Du>{1}/{4}$ or $\Dd>{1}/{4}$. Actually, the helicity 0 component has a smaller range of convergence: $0 < \Du,\Dd  < {1}/{4}$. 

In the range of convergence, the ratio among the two helicities agrees with the ratio of the corresponding coefficients in Eq.~(\ref{vector_2-point_fourier}). In terms of $\Delta_v$ the range of convergence is $1 < \Delta_v < \frac{3}{2}$. 
For $\Delta_v > \frac{3}{2}$, the two helicity components diverge in the same way keeping their ratio constant. 
This hints  that this divergence is not physical, the 2-point function should be renormalized.
For $0<\Delta_v<1$, the helicity zero component is still positive, however, it diverges.

\paragraph{Spin 2 case}
Let us verify that the 2-point correlation function of $S_{ij}$ defined in Eq.~\eqref{O2} is of the form \eqref{S2real}. Since we are going to contract the operator with null polarization vectors $\vec \epsilon$ and $\vec{\tilde \epsilon}$, the terms proportional $\delta_{ij}$ can be dropped:
\begin{eqnarray}
 \langle{S^{ij}(\vec x) S^{kl}(0)}\rangle &=& \Delta^2 \langle{\phi\prt _i\prt _j\phi(\vec x)\ \phi\prt _k\prt _l\phi(0)}\rangle
-\Delta(\Delta+1) \langle{\phi\prt _i\prt _j\phi(\vec x)\ \prt _k\phi\prt_l\phi(0)}\rangle \nonumber\\
&&-\Delta(\Delta+1) \langle{\prt_i\phi\prt_j\phi(\vec x)\ \phi\prt_k\prt_l\phi(0)}\rangle
+(\Delta+1)^2 \langle{\prt_i\phi\prt_j\phi(\vec x)\ \prt_k\phi\prt _l\phi(0)}\rangle.
\end{eqnarray}
In the Gaussian approximation, this can be evaluated by taking products of derivatives of
\be
\langle{\phi(\vec x)\phi(0)}\rangle = \frac{1}{| x|^{2\Delta}}.
\ee
For instance 
\be
\langle{\phi\prt_i\prt_j\phi(\vec x)\ \phi\prt_k\prt_l\phi(\vec y)}\rangle=\langle{\phi(\vec x)\ \phi(\vec y)}\rangle\langle{\prt_i\prt_j\phi(\vec x)\ \prt_k\prt_l\phi(\vec y)}\rangle
+\langle{\phi(\vec x)\ \prt_k\prt_l\phi(\vec y)}\rangle\langle{\prt_i\prt_j\phi(\vec x)\ \phi(\vec y)}\rangle.
\ee
Neglecting terms proportional to $\delta_{ij}$ or $\delta_{kl}$ we get for this term
\begin{eqnarray}
\langle{\phi\prt_i\prt_j\phi(\vec x)\ \phi\prt_k\prt_l\phi(0)}\rangle&=&\frac{1}{| x|^{2(2\Delta+2)}}
\left[ 4\Delta(\Delta+1)(\delta^{ik}\delta^{jl}+\delta^{il}\delta^{jk}) 
 -8 \Delta(\Delta+1)(\Delta+2)(\delta^{ik}\hat x^j\hat x^l +\text{3 Perm.})\right. \nonumber \\
&&\left.+16 \Delta(\Delta+1)[(\Delta+2)(\Delta+3)+\Delta(\Delta+1)]\hat x^i\hat x^j\hat x^k\hat x^l\right].  
\end{eqnarray}
The other two contributions are given by
\be
\langle{\phi\prt_i\prt_j\phi(\vec x)\ \prt_k\phi\prt_l\phi(0)}\rangle=
\frac{1}{|x|^{2(2\Delta+2)}}
\left[ -8 \Delta^2(\Delta+1)(\delta^{ik}\hat x^j\hat x^l +\text{3 Perm.})
+32 \Delta^2(\Delta+1)(\Delta+2)\hat x^i\hat x^j\hat x^k\hat x^l\right]. \nonumber
\ee
\be
\langle{\prt_i\phi\prt_j\phi(\vec x)\ \prt_k\phi\prt_l\phi(0)}=
\frac{1}{|x|^{2(2\Delta+2)}}
\left[ 4\Delta^2(\delta^{ik}-2(\Delta+1)\hat x^i\hat x^k)(\delta^{jl}-2(\Delta+1)\hat x^j\hat x^l)+(k\leftrightarrow l)\right].
\ee
Summing them up we get 
\be
\langle{S^{ij}(\vec x) S^{kl}(0)}\rangle=\frac{4\Delta^2(2\Delta+1)}{|x|^{2(2\Delta+2)}}
(\delta^{ik}-2\hat x^i\hat x^k)(\delta^{jl}-2\hat x^j\hat x^l)+(k\leftrightarrow l)
\ee
plus terms that vanish when contracted with null polarization vectors. 

\section{\label{sec:waveuniverse}Wave function of the Universe}

An alternative way to compute expectation values of the primordial perturbations consists in considering the wave function of the Universe. In this approach cosmological expectation values are casted in terms of the functional $\psi[\chi]$, with $\chi$ standing for a generic perturbation. The perturbations produced during inflation are known to be approximately Gaussian. This allows the wave function to be written as a power series expansion of the form,
\be
￼￼\Psi[\chi(\vec x)]= \exp \left[􏰋−\frac{1}{2}􏰕 \! \int \!\! d^3x\,d^3y\ \chi(\vec x)\chi(\vec y) \, \langle O(\vec x)O(\vec y)\rangle + \dots \right], 
\ee
where the coefficient $\langle O(x)O(y) \rangle$ determines the two-point correlators. 

If dS isometries are respected by the fluctuations then the coefficient functions will transform under the $SO(4,1)$ symmetries like correlation functions of appropriate operators in a Euclidean CFT. This means that also the functions $\langle O( \vec x)O(\vec y) \rangle$ must satisfies Eq.~\eqref{S2real}. We focus just on the tensor perturbation $S_{ij}$, hence, in Fourier space,
\begin{equation}
\Psi[S_{ij}] \sim \mbox{exp}\left[ -\frac{1}{2} \int \!\! \frac{\mathrm{d}^3 k}{(2\pi)^3} \, S_{ij} S_{kl} \langle \sigma_{ij}( \vec k) \sigma_{kl}(\!-\!\vec k) \rangle'  \right]\,.
\end{equation}
In the above expression, the function $\langle \epsilon^2. \sigma_{ \vec k} \tilde \epsilon^2.\sigma_{\!-\!\vec k} \rangle'$ satisfies Eq.~\eqref{spin_2_2-point_fourier} with $\Delta= \Delta_+$. Let us verify this by deriving the two point function for the spin-2 field $S_{ij}$ assuming that $\langle \epsilon^2. \sigma_{ \vec k} \tilde \epsilon^2.\sigma_{\!-\!\vec k} \rangle'$ satisfies Eq.~\eqref{spin_2_2-point_fourier} with the larger dimension $\Delta\equiv \Delta_+$.
First we need to derive $\langle{\sigma_{ij}\sigma_{kl}}\rangle$ from Eq.~\eqref{spin_2_2-point_fourier} which can be parameterized as
\be\label{OO}
\langle{\epsilon^2. \sigma \, \tilde\epsilon^2. \sigma}\rangle\propto \tilde a e^{2i\theta} + \tilde b e^{i\theta} +\tilde c
+ \tilde b e^{-i\theta}+\tilde a e^{-2i\theta} 
\ee
with 
\be
\tilde a =1,\quad \tilde b = 4\frac{3-\Delta}{\Delta},\quad \tilde c = 6\frac{(3-\Delta)(2-\Delta)}{(\Delta-1)\Delta}.
\ee
Eq.~\eqref{OO} can be written as
\be
a (\epsilon \!\cdot\! \tilde\epsilon)^2 + b(\epsilon \!\cdot\! \tilde\epsilon)+c
\ee
with
\be
a = 4\tilde a ,\quad b = 2\tilde b -8\tilde a,\quad c=\tilde c +2\tilde a -2\tilde b.
\ee
Therefore, we have
\begin{eqnarray}\label{OO2}
\langle{\sigma_{ij}\sigma_{kl}}\rangle&\propto& \frac{1}{2} a (\delta_{ik}\delta_{jl}+\delta_{il}\delta_{jk})
+\frac{1}{4}b(\delta_{ik}\hat k_j\hat k_l+\delta_{il}\hat k_j\hat k_k+\delta_{jk}\hat k_i\hat k_l+\delta_{jl}\hat k_i\hat k_k)
+c\hat k_i\hat k_j\hat k_k\hat k_l \nonumber \\
&&+d \, \delta_{ij}\delta_{kl}+e(\delta_{ij}\hat k_l\hat k_k +\delta_{kl}\hat k_i\hat k_j ).
\end{eqnarray}
The coefficients $d,e$ are determined in terms of the other three by the tracelessness condition:
\be
e=-\frac{1}{3}(b+c),\qquad d=-\frac{1}{3}a+\frac{1}{9}(b+c).
\ee
To derive $\langle{S_{ij}S_{kl}}\rangle$ we should invert this matrix:
\begin{eqnarray}\label{ss}
\langle{S_{ij}S_{kl}}\rangle &\propto & A (\delta_{ik}\delta_{jl}+\delta_{il}\delta_{jk})
+B(\delta_{ik}\hat k_j\hat k_l+\delta_{il}\hat k_j\hat k_k+\delta_{jk}\hat k_i\hat k_l+\delta_{jl}\hat k_i\hat k_k)
+C\hat k_i\hat k_j\hat k_k\hat k_l \nonumber \\
&&+D \delta_{ij}\delta_{kl}+e(\delta_{ij}\hat k_l\hat k_k +\delta_{kl}\hat k_i\hat k_j ),
\end{eqnarray}
where tracelessness condition fixes
\be
E=-\frac{1}{3}(4B+C),\qquad D=-\frac{2}{3}A-\frac{1}{9}(4B+C),
\ee
and the other three coefficients are~\footnote{A simplification arises by noting that Eq.~\eqref{OO2} is traceless.}
\be
B=\frac{-b}{(2a+b)}A,\qquad C = \frac{-2cA -(4c+2b+4e)B}{a+b+c+e}.
\ee
The correlator Eq.~\eqref{ss} can be contracted with polarization vectors to be brought back to the form Eq.~\eqref{OO}, with 
\be
\tilde A = \frac{1}{2} A,\quad \tilde B = 4(\frac{B}{A}+1) \tilde A,\quad \tilde C = C+3A+4B.
\ee
One obtains
\bea
\frac{\tilde B}{\tilde A} &=& 4\frac{\Delta}{3-\Delta}=4\frac{3-\Delta_-}{\Delta_-}, \nonumber \\
\frac{\tilde C}{\tilde A} &=& 6\frac{\Delta(\Delta-1)}{(3-\Delta)(2-\Delta)}
=6\frac{(3-\Delta_-)(2-\Delta_-)}{\Delta_-(\Delta_- -1)},
\eea
which are the same as Eq.~\eqref{OO} with $\Delta\to \Delta_-$.

\def\thefootnote{\fnsymbol{footnote}}


\footnotesize
\parskip 0pt

\begin{thebibliography}{99}

\bibitem{Weinberg:1964ew} 
  S.~Weinberg,
  ``Photons and Gravitons in S Matrix Theory: Derivation of Charge Conservation and Equality of Gravitational and Inertial Mass,''
  Phys.\ Rev.\  {\bf 135}, B1049 (1964).
  
\bibitem{Weinberg:1965rz} 
  S.~Weinberg,
  ``Photons and gravitons in perturbation theory: Derivation of Maxwell's and Einstein's equations,''
  Phys.\ Rev.\  {\bf 138}, B988 (1965).
  
\bibitem{deRham:2014zqa} 
  C.~de Rham,
  ``Massive Gravity,''
  Living Rev.\ Rel.\  {\bf 17}, 7 (2014)
  [arXiv:1401.4173 [hep-th]].

\bibitem{Creminelli:2014wna} 
  P.~Creminelli, J.~Gleyzes, J.~Nore\~na and F.~Vernizzi,
  ``Resilience of the standard predictions for primordial tensor modes,''
  Phys.\ Rev.\ Lett.\  {\bf 113}, no. 23, 231301 (2014)
  [arXiv:1407.8439 [astro-ph.CO]].

\bibitem{Kleban:2015daa} 
  M.~Kleban, M.~Mirbabayi and M.~Porrati,
  ``Effective Planck Mass and the Scale of Inflation,''
  JCAP {\bf 1601}, no. 01, 017 (2016)
  [arXiv:1508.01527 [hep-th]].


\bibitem{Cheung:2007st} 
  C.~Cheung, P.~Creminelli, A.~L.~Fitzpatrick, J.~Kaplan and L.~Senatore,
  ``The Effective Field Theory of Inflation,''
  JHEP {\bf 0803}, 014 (2008)
  [arXiv:0709.0293 [hep-th]].
      
\bibitem{Maldacena:2002vr} 
  J.~M.~Maldacena,
  ``Non-Gaussian features of primordial fluctuations in single field inflationary models,''
  JHEP {\bf 0305}, 013 (2003)
  [astro-ph/0210603].
  
\bibitem{Creminelli:2004yq} 
  P.~Creminelli and M.~Zaldarriaga,
  ``Single field consistency relation for the 3-point function,''
  JCAP {\bf 0410}, 006 (2004)
  [astro-ph/0407059].
  
\bibitem{Dimastrogiovanni:2015pla} 
  E.~Dimastrogiovanni, M.~Fasiello and M.~Kamionkowski,
  ``Imprints of Massive Primordial Fields on Large-Scale Structure,''
  JCAP {\bf 1602}, 017 (2016)
  [arXiv:1504.05993 [astro-ph.CO]].
  
\bibitem{Weinberg} 
  S.~Weinberg,
  ``Adiabatic modes in cosmology,''
  Phys.\ Rev.\ D {\bf 67}, 123504 (2003)
  [astro-ph/0302326].
  
\bibitem{Creminelli:2013mca} 
  P.~Creminelli, J.~Noreña, M.~Simonović and F.~Vernizzi,
  ``Single-Field Consistency Relations of Large Scale Structure,''
  JCAP {\bf 1312}, 025 (2013)
  [arXiv:1309.3557 [astro-ph.CO]].
  
\bibitem{Bartolo:2013msa} 
  N.~Bartolo, S.~Matarrese, M.~Peloso and A.~Ricciardone,
  ``Anisotropy in solid inflation,''
  JCAP {\bf 1308}, 022 (2013)
  [arXiv:1306.4160 [astro-ph.CO]].
 
\bibitem{Akhshik:2014gja} 
  M.~Akhshik, R.~Emami, H.~Firouzjahi and Y.~Wang,
  ``Statistical Anisotropies in Gravitational Waves in Solid Inflation,''
  JCAP {\bf 1409}, 012 (2014)
  [arXiv:1405.4179 [astro-ph.CO]].
 
\bibitem{Higuchi:1986py} 
  A.~Higuchi,
  ``Forbidden Mass Range for Spin-2 Field Theory in De Sitter Space-time,''
  Nucl.\ Phys.\ B {\bf 282}, 397 (1987).
  
\bibitem{Wald} 
  R.~M.~Wald,
  ``Asymptotic behavior of homogeneous cosmological models in the presence of a positive cosmological constant,''
  Phys.\ Rev.\ D {\bf 28}, 2118 (1983).
  
\bibitem{Gruzinov:2004ty} 
  A.~Gruzinov,
  ``Elastic inflation,''
  Phys.\ Rev.\ D {\bf 70}, 063518 (2004)
  [astro-ph/0404548].
  
\bibitem{Golovnev:2008cf} 
  A.~Golovnev, V.~Mukhanov and V.~Vanchurin,
  ``Vector Inflation,''
  JCAP {\bf 0806}, 009 (2008)
  [arXiv:0802.2068 [astro-ph]].
  
\bibitem{Maleknejad:2011sq} 
  A.~Maleknejad and M.~M.~Sheikh-Jabbari,
  ``Non-Abelian Gauge Field Inflation,''
  Phys.\ Rev.\ D {\bf 84}, 043515 (2011)
  [arXiv:1102.1932 [hep-ph]].

  
\bibitem{Endlich:2012pz} 
  S.~Endlich, A.~Nicolis and J.~Wang,
  ``Solid Inflation,''
  JCAP {\bf 1310}, 011 (2013)
  [arXiv:1210.0569 [hep-th]].
    
\bibitem{Adshead:2013qp} 
  P.~Adshead, E.~Martinec and M.~Wyman,
  ``Gauge fields and inflation: Chiral gravitational waves, fluctuations, and the Lyth bound,''
  Phys.\ Rev.\ D {\bf 88}, no. 2, 021302 (2013)
  [arXiv:1301.2598 [hep-th]].
  
\bibitem{Adshead:2013nka} 
  P.~Adshead, E.~Martinec and M.~Wyman,
  ``Perturbations in Chromo-Natural Inflation,''
  JHEP {\bf 1309}, 087 (2013)
  [arXiv:1305.2930 [hep-th]].
  
\bibitem{Cannone:2014uqa} 
  D.~Cannone, G.~Tasinato and D.~Wands,
  ``Generalised tensor fluctuations and inflation,''
  JCAP {\bf 1501}, no. 01, 029 (2015)
  [arXiv:1409.6568 [astro-ph.CO]].
    
\bibitem{Bartolo:2015qvr} 
  N.~Bartolo, D.~Cannone, A.~Ricciardone and G.~Tasinato,
  ``Distinctive signatures of space-time diffeomorphism breaking in EFT of inflation,''
  JCAP {\bf 1603}, no. 03, 044 (2016)
  [arXiv:1511.07414 [astro-ph.CO]].
  
\bibitem{Meerburg:2016ecv} 
  P.~D.~Meerburg, J.~Meyers, A.~van Engelen and Y.~Ali-Haïmoud,
  ``On CMB B-Mode Non-Gaussianity,''
  arXiv:1603.02243 [astro-ph.CO].
  
\bibitem{Hinterbichler} 
  K.~Hinterbichler, L.~Hui and J.~Khoury,
  ``An Infinite Set of Ward Identities for Adiabatic Modes in Cosmology,''
  arXiv:1304.5527 [hep-th].

\bibitem{double} 
  M.~Mirbabayi and M.~Zaldarriaga,
  ``Double Soft Limits of Cosmological Correlations,''
  JCAP {\bf 1503}, no. 03, 025 (2015)
  [arXiv:1409.6317 [hep-th]].
  
\bibitem{Lyth:2001nq} 
  D.~H.~Lyth and D.~Wands,
  ``Generating the curvature perturbation without an inflaton,''
  Phys.\ Lett.\ B {\bf 524}, 5 (2002)
  [hep-ph/0110002].
    
\bibitem{Chen:2009we} 
  X.~Chen and Y.~Wang,
  ``Large non-Gaussianities with Intermediate Shapes from Quasi-Single Field Inflation,''
  Phys.\ Rev.\ D {\bf 81}, 063511 (2010)
  [arXiv:0909.0496 [astro-ph.CO]].
  
\bibitem{Namjoo:2012aa} 
  M.~H.~Namjoo, H.~Firouzjahi and M.~Sasaki,
  ``Violation of non-Gaussianity consistency relation in a single field inflationary model,''
  Europhys.\ Lett.\  {\bf 101}, 39001 (2013)
  [arXiv:1210.3692 [astro-ph.CO]].
  
  \bibitem{Flauger} 
  R.~Flauger, D.~Green and R.~A.~Porto,
  ``On squeezed limits in single-field inflation.  Part I,''
  JCAP {\bf 1308}, 032 (2013)
  [arXiv:1303.1430 [hep-th]].
  
\bibitem{Cook:2013xea} 
  J.~L.~Cook and L.~Sorbo,
  JCAP {\bf 1311}, 047 (2013)
  doi:10.1088/1475-7516/2013/11/047
  [arXiv:1307.7077 [astro-ph.CO]].
  
\bibitem{Mukohyama:2014gba} 
  S.~Mukohyama, R.~Namba, M.~Peloso and G.~Shiu,
  ``Blue Tensor Spectrum from Particle Production during Inflation,''
  JCAP {\bf 1408}, 036 (2014)
  [arXiv:1405.0346 [astro-ph.CO]].
  
\bibitem{Ferreira:2014zia} 
  R.~Z.~Ferreira and M.~S.~Sloth,
  ``Universal Constraints on Axions from Inflation,''
  JHEP {\bf 1412}, 139 (2014)
  [arXiv:1409.5799 [hep-ph]].
  
\bibitem{Ferreira:2015omg} 
  R.~Z.~Ferreira, J.~Ganc, J.~Noreña and M.~S.~Sloth,
  ``On the validity of the perturbative description of axions during inflation,''
  JCAP {\bf 1604}, no. 04, 039 (2016)
  [arXiv:1512.06116 [astro-ph.CO]].
    
\bibitem{Mirbabayi:2014jqa} 
  M.~Mirbabayi, L.~Senatore, E.~Silverstein and M.~Zaldarriaga,
  ``Gravitational Waves and the Scale of Inflation,''
  Phys.\ Rev.\ D {\bf 91}, 063518 (2015)
  [arXiv:1412.0665 [hep-th]].

\bibitem{Flauger_axion} 
  R.~Flauger, L.~McAllister, E.~Pajer, A.~Westphal and G.~Xu,
  ``Oscillations in the CMB from Axion Monodromy Inflation,''
  JCAP {\bf 1006}, 009 (2010)
  [arXiv:0907.2916 [hep-th]].
  
\bibitem{Maleknejad} 
  A.~Maleknejad and M.~M.~Sheikh-Jabbari,
  ``Revisiting Cosmic No-Hair Theorem for Inflationary Settings,''
  Phys.\ Rev.\ D {\bf 85}, 123508 (2012)
  [arXiv:1203.0219 [hep-th]].

\bibitem{Ratra:1991bn} 
  B.~Ratra,
  ``Cosmological 'seed' magnetic field from inflation,''
  Astrophys.\ J.\  {\bf 391}, L1 (1992).

\bibitem{Watanabe:2009ct} 
  M.~a.~Watanabe, S.~Kanno and J.~Soda,
  ``Inflationary Universe with Anisotropic Hair,''
  Phys.\ Rev.\ Lett.\  {\bf 102}, 191302 (2009)
  doi:10.1103/PhysRevLett.102.191302
  [arXiv:0902.2833 [hep-th]].

\bibitem{Endlich:2013jia} 
  S.~Endlich, B.~Horn, A.~Nicolis and J.~Wang,
  ``Squeezed limit of the solid inflation three-point function,''
  Phys.\ Rev.\ D {\bf 90}, no. 6, 063506 (2014)
  [arXiv:1307.8114 [hep-th]].
    
\bibitem{Fasiello:2012rw} 
  M.~Fasiello and A.~J.~Tolley,
  ``Cosmological perturbations in Massive Gravity and the Higuchi bound,''
  JCAP {\bf 1211}, 035 (2012)
  [arXiv:1206.3852 [hep-th]].
  
\bibitem{squeezed} 
  M.~Mirbabayi and M.~Simonovi\'c,
  ``Effective Theory of Squeezed Correlation Functions,''
  JCAP {\bf 1603}, no. 03, 056 (2016)
  [arXiv:1507.04755 [hep-th]].
  
\bibitem{Arkani-Hamed:2015bza} 
  N.~Arkani-Hamed and J.~Maldacena,
  ``Cosmological Collider Physics,''
  arXiv:1503.08043 [hep-th].
  
\bibitem{Pajer:2013ana} 
  E.~Pajer, F.~Schmidt and M.~Zaldarriaga,
  ``The Observed Squeezed Limit of Cosmological Three-Point Functions,''
  Phys.\ Rev.\ D {\bf 88}, no. 8, 083502 (2013)
  [arXiv:1305.0824 [astro-ph.CO]].
  
\bibitem{Masui:2010cz} 
  K.~W.~Masui and U.~L.~Pen,
  ``Primordial gravity wave fossils and their use in testing inflation,''
  Phys.\ Rev.\ Lett.\  {\bf 105}, 161302 (2010)
  [arXiv:1006.4181 [astro-ph.CO]].
 
\bibitem{Giddings:2011zd} 
  S.~B.~Giddings and M.~S.~Sloth,
 ``Cosmological observables, IR growth of fluctuations, and scale-dependent anisotropies,''
  Phys.\ Rev.\ D {\bf 84}, 063528 (2011)
  doi:10.1103/PhysRevD.84.063528
  [arXiv:1104.0002 [hep-th]].
 
\bibitem{Schmidt:2012nw} 
  F.~Schmidt and D.~Jeong,
  ``Large-Scale Structure with Gravitational Waves II: Shear,''
  Phys.\ Rev.\ D {\bf 86}, 083513 (2012)
  [arXiv:1205.1514 [astro-ph.CO]].
  
\bibitem{Dai:2013kra} 
  L.~Dai, D.~Jeong and M.~Kamionkowski,
  ``Anisotropic imprint of long-wavelength tensor perturbations on cosmic structure,''
  Phys.\ Rev.\ D {\bf 88}, no. 4, 043507 (2013)
  [arXiv:1306.3985 [astro-ph.CO]].
  
\bibitem{Schmidt:2013gwa} 
  F.~Schmidt, E.~Pajer and M.~Zaldarriaga,
  ``Large-Scale Structure and Gravitational Waves III: Tidal Effects,''
  Phys.\ Rev.\ D {\bf 89}, no. 8, 083507 (2014)
  [arXiv:1312.5616 [astro-ph.CO]].
  
\bibitem{Ade:2015xua} 
  P.~A.~R.~Ade {\it et al.} [Planck Collaboration],
  ``Planck 2015 results. XIII. Cosmological parameters,''
  arXiv:1502.01589 [astro-ph.CO].
  
\bibitem{Brahma:2013rua} 
  S.~Brahma, E.~Nelson and S.~Shandera,
  ``Fossilized Gravitational Wave Relic and Primordial Clocks,''
  Phys.\ Rev.\ D {\bf 89}, no. 2, 023507 (2014)
  [arXiv:1310.0471 [astro-ph.CO]].
  
  \cite{Dimastrogiovanni:2014ina}
\bibitem{Dimastrogiovanni:2014ina} 
  E.~Dimastrogiovanni, M.~Fasiello, D.~Jeong and M.~Kamionkowski,
  ``Inflationary tensor fossils in large-scale structure,''
  JCAP {\bf 1412}, 050 (2014)
  [arXiv:1407.8204 [astro-ph.CO]].
  
\bibitem{Akhshik:2014bla} 
  M.~Akhshik,
  ``Clustering Fossils in Solid Inflation,''
  JCAP {\bf 1505}, no. 05, 043 (2015)
  [arXiv:1409.3004 [astro-ph.CO]].

\bibitem{Emami:2015uva} 
  R.~Emami and H.~Firouzjahi,
  ``Clustering Fossil from Primordial Gravitational Waves in Anisotropic Inflation,''
  JCAP {\bf 1510}, no. 10, 043 (2015)
  [arXiv:1506.00958 [astro-ph.CO]].
  
\bibitem{Ade:2015hxq} 
  P.~A.~R.~Ade {\it et al.} [Planck Collaboration],
  ``Planck 2015 results. XVI. Isotropy and statistics of the CMB,''
  arXiv:1506.07135 [astro-ph.CO].
     
\bibitem{Bartolo:2014xfa} 
  N.~Bartolo, M.~Peloso, A.~Ricciardone and C.~Unal,
  ``The expected anisotropy in solid inflation,''
  JCAP {\bf 1411}, no. 11, 009 (2014)
  [arXiv:1407.8053 [astro-ph.CO]].
  
\bibitem{Shiraishi:2016omb} 
  M.~Shiraishi, J.~B.~Mu\~noz, M.~Kamionkowski and A.~Raccanelli,
  ``Violation of statistical isotropy and homogeneity in the 21-cm power spectrum,''
  arXiv:1603.01206 [astro-ph.CO].
    
\bibitem{Ackerman:2007nb} 
  L.~Ackerman, S.~M.~Carroll and M.~B.~Wise,
  ``Imprints of a Primordial Preferred Direction on the Microwave Background,''
  Phys.\ Rev.\ D {\bf 75}, 083502 (2007)
  Erratum: [Phys.\ Rev.\ D {\bf 80}, 069901 (2009)]
  [astro-ph/0701357].
  
\bibitem{Watanabe:2010fh} 
  M.~a.~Watanabe, S.~Kanno and J.~Soda,
  ``The Nature of Primordial Fluctuations from Anisotropic Inflation,''
  Prog.\ Theor.\ Phys.\  {\bf 123}, 1041 (2010)
  doi:10.1143/PTP.123.1041
  [arXiv:1003.0056 [astro-ph.CO]].
  
\bibitem{Bartolo:2012sd} 
  N.~Bartolo, S.~Matarrese, M.~Peloso and A.~Ricciardone,
  ``Anisotropic power spectrum and bispectrum in the $f(\phi)F^2$ mechanism,''
  Phys.\ Rev.\ D {\bf 87}, no. 2, 023504 (2013)
  [arXiv:1210.3257 [astro-ph.CO]].

  
\bibitem{Doroshkevich} 
  A.~G.~Doroshkevich,  ``Spatial structure of perturbations and origin of galactic rotation in fluctuation theory''
  Astrofizika, vol.~6, no.~4, 581 (1970)
  
\bibitem{Seery:2008ax} 
  D.~Seery, M.~S.~Sloth and F.~Vernizzi,
  ``Inflationary trispectrum from graviton exchange,''
  JCAP {\bf 0903}, 018 (2009)
  [arXiv:0811.3934 [astro-ph]].
  
\bibitem{Byrnes:2006vq} 
  C.~T.~Byrnes, M.~Sasaki and D.~Wands,
  ``The primordial trispectrum from inflation,''
  Phys.\ Rev.\ D {\bf 74}, 123519 (2006)
  [astro-ph/0611075].
  
\bibitem{Ade:2013ydc} 
  P.~A.~R.~Ade {\it et al.} [Planck Collaboration],
  ``Planck 2013 Results. XXIV. Constraints on primordial non-Gaussianity,''
  Astron.\ Astrophys.\  {\bf 571}, A24 (2014)
  [arXiv:1303.5084 [astro-ph.CO]].
  
\bibitem{Jeong:2012df}
  D.~Jeong and M.~Kamionkowski,
  ``Clustering Fossils from the Early Universe,''
  Phys.\ Rev.\ Lett.\  {\bf 108} (2012) 251301
  [arXiv:1203.0302 [astro-ph.CO]].
   
%
    
\bibitem{Blas:2009my} 
  D.~Blas, D.~Comelli, F.~Nesti and L.~Pilo,
  ``Lorentz Breaking Massive Gravity in Curved Space,''
  Phys.\ Rev.\ D {\bf 80}, 044025 (2009)
  doi:10.1103/PhysRevD.80.044025
  [arXiv:0905.1699 [hep-th]].
  
\bibitem{local} 
  P.~Creminelli, L.~Senatore and M.~Zaldarriaga,
  ``Estimators for local non-Gaussianities,''
  JCAP {\bf 0703}, 019 (2007)
  doi:10.1088/1475-7516/2007/03/019
  [astro-ph/0606001].


 

\end{thebibliography}
\end{document}